\documentclass[letterpaper,twocolumn,10pt]{article}
\usepackage[latin1]{inputenc}
\usepackage{graphicx}
\usepackage{flushend}
\usepackage{authblk}
\usepackage[english]{babel}
\usepackage{multirow, array} 
\usepackage{nicefrac}
\usepackage{multicol}
\usepackage{xcolor}
\usepackage[backend=biber,style=numeric-comp,sorting=none]{biblatex}
\usepackage{amssymb}

\usepackage{biblatex} 
\date{}

\begin{document}

\title{Bright squeezed vacuum for two-photon spectroscopy: simultaneously high resolution in time and frequency, space and wavevector.}

\author[1,2,*]{Paula Cutipa}
\author[1,2]{Maria V. Chekhova}

\affil[1]{Max-Planck Institute for the Science of Light, Staudtstra{\ss}e 2, Erlangen D-91058, Germany}
\affil[2]{University of Erlangen-Nuremberg, Staudtstra{\ss}e 7/B2, 91058 Erlangen, Germany}

\affil[*]{Corresponding author: paula.cutipa@mpl.mpg.de}

\twocolumn[
\begin{@twocolumnfalse}
\maketitle
\vspace*{-1cm}
\begin{center}\rule{0.9\textwidth}{0.1mm} \end{center}
\normalsize Entangled photons offer two advantages for two-photon absorption spectroscopy. One of them, the linear scaling of two-photon absorption rate with the input photon flux, is only valid at very low photon fluxes and is therefore impractical. The other is the overcoming of the classical constraints for simultaneous resolution in time-frequency and in space-wavevector. Here we consider bright squeezed vacuum (BSV) as an alternative to entangled photons. The efficiency increase it offers in comparison with coherent light is modest, but it does not depend on the photon flux. Moreover, and this is what we show in this work, BSV also provides simultaneously high resolution in time and frequency, and in space and wavevector. In our experiment, we measure the widths of the second-order correlation functions in space, time, frequency, and angle, and demonstrate the violation of the constraint given by the Fourier transformation, also known as the Mancini criterion of entanglement.
\begin{center}\rule{0.9\textwidth}{0.1mm} \end{center}
\vspace*{0.5cm}
\end{@twocolumnfalse}
]

\maketitle

Entangled photons play an important role in different applications such as quantum imaging,  microscopy, spectroscopy, and metrology~\cite{clark2021special,dorfman2016nonlinear}. In particular, energy-time and position-momentum entanglement permits the enhancement of applications based on two-photon absorption (TPA), such as TPA spectroscopy for complex sample analysis or cellular imaging ~\cite{dayan2005nonlinear,schlawin2018entangled,leontemperature,mikhaylov2020comprehensive,villabona2020measurements,tabakaev2021energy}.
 One of the advantages of using entangled photons is that it improves the TPA rate, which then scales with the photon flux linearly, and not quadratically, as in the case of coherent light~\cite{klyshko1982transverse,javanainen1990linear}. This happens because the TPA rate is proportional to the second-order correlation function (CF) of the incident light \cite{glauber1963quantum} at zero delay and displacement, $g^{(2)}(0)$, which for a weak flux of photon pairs scales inversely with the mean photon number \cite{klyshko1982transverse}. This enables the analysis of biological samples at low light intensities, avoiding damage to the sample \cite{dayan2005nonlinear,javanainen1990linear}. However, this linear dependence is only valid at very low photon fluxes, where it can be hardly used in practice~\cite{mikhaylov2020comprehensive,raymer2021large}. 

Another advantage of using photon pairs for TPA spectroscopy is the simultaneously high and independent resolution in time ($\delta \tau$) and frequency ($\delta \omega$) they provide~\cite{schlawin2018entangled}. Indeed, photons entangled in time and frequency violate the classical relation $\delta\omega \delta \tau \geq 2\pi$, also known as the Mancini separability bound~\cite{mancini2002entangling}. The same situation is valid for the transverse wavevector ($\delta k$) and space ($\delta \xi$) \cite{howell2004realization}. This enables  simultaneously high resolution in time, space, frequency, and wavevector.

Entangled photons are usually generated by spontaneous parametric down-conversion (PDC). When strongly pumped, PDC produces a brighter state of light, bright squeezed vacuum (BSV) \cite{iskhakov2009generation,spasibko2012spectral,brambilla2012disclosing,allevi2014coherence}. For this state, $g^{(2)}(0)=3$, which provides only threefold enhancement of two-photon absorption compared to the case of coherent light excitation \cite{spasibko2017multiphoton}, but this enhancement is independent of the photon flux. Meanwhile, the mean number of photons per one mode of BSV, $N=\sinh^2(G)$, where $G$ is called the parametric gain, can be as high as for laser sources. This makes BSV a prospective source for practical use in multiphoton absorption. Below we show that the second advantage of entangled photons, the simultaneously high resolution in time, space, frequency, and wavevector, is also valid for BSV.

The time, frequency, space, and wavevector resolution for two-photon transitions is determined by the second-order CF. The CF, in the general case, is a nonfactorable function of four variables: two time moments and two space positions (in one dimension, for simplicity), $g^{(2)}(t,t';x,x')$~\cite{brambilla2012disclosing,caspani2010tailoring,gatti2009x}. Alternatively, the CF can be considered as a function of frequencies and transverse wavevectors, $g^{(2)}(\omega,\omega';k,k')$ \cite{finger2017characterization}.

At low parametric gain, $G\ll1$, $g^{(2)}(0)\approx 1/G^2$ and is very high. 
The resolution provided by entangled photons in two-photon spectroscopy is given by its widths with respect to the corresponding variables: for instance, $\delta\tau$ is the width of $g^{(2)}(t,t';x,x)$ at fixed $t',x$, and similarly with the other variables. 

At high gain, $G\gtrsim1$, photon pairs overlap, and their accidental coincidences, as well as thermal correlations of photons within one mode, start to play a role. Then, the value of $g^{(2)}$ reduces \cite{klyshko1982transverse,allevi2014coherence}, but its widths in time  $\delta\tau$, space $\delta\xi$, frequency $\delta\omega$, and transverse wavevector $\delta k$, further called {\it correlation widths}, do not change much.

In this work we measure all these four widths. To measure $\delta\tau$ and $\delta\xi$, we build an ultra-fast autocorrelator based on sum-frequency generation (SFG) ~\cite{dayan2004two,brambilla2012disclosing} in a non-phase-matched crystal. By using type-II SFG we get rid of the first-order interference fringes~\cite{boitier2013two}. The values of $\delta\omega$ and $\delta k$  we obtain from the measurement of the covariance of BSV intensity~\cite{spasibko2012spectral,sharapova2020properties}.

\begin{figure}[htbt] 
\centering
\includegraphics[width=0.9\linewidth]{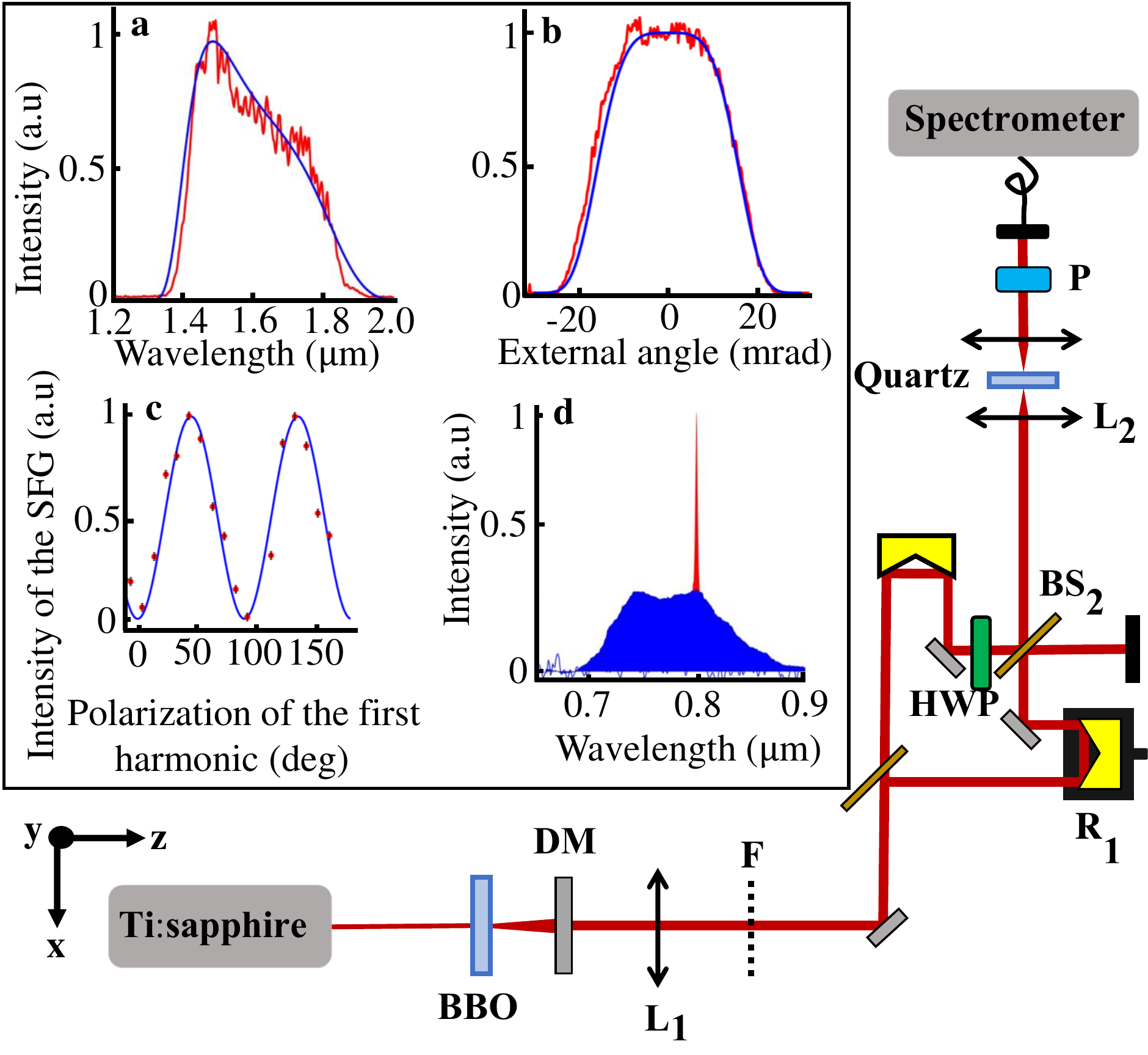}
\caption{Experimental setup for measuring the second-order CF. We produced BSV in the BBO crystal whose output face was imaged on beam splitter $BS_2$ by lens $L_1$ and then on the quartz crystal by lens $L_2$. In the autocorrelator, retroreflector $R_1$ scanned the time delay and beamsplitter $BS_2$ scanned the spatial displacement. In one arm, a HWP changed the polarisation, so that type-II SFG was generated in quartz at the output and registered, after polarizer P, by a spectrometer. The insets show the wavelength and angular spectra of BSV emitted via collinear degenerate PDC (a, b), the  dependence of the SFG signal on the input polarization with one arm blocked (c) and the SFG spectrum (d).}
\label{fig:Experimental-setup}
\end{figure}

The experimental setup is  displayed in Fig. \ref{fig:Experimental-setup}. We generated BSV with an amplified Ti-sapphire laser (wavelength $800$ nm, pulse duration $1.6$ ps, repetition rate $5$ kHz, mean power $1W$, waist $0.8$ mm) in a $10$ mm $\beta$-barium-borate (BBO) crystal cut for type-I phase matching. The experiment was run at gain $G=10$, both under collinear (the pump wavevector at $19.87^{o}$ to the optic axis) and non-collinear (the pump wavevector at $19.93^{o}$ to the optic axis) phase matching. In both cases, the angular-frequency spectrum of BSV is bounded, however, it is factorable only in the collinear configuration ~\cite{spasibko2016ring}.  
The optic axis of the BBO crystal was in the y-z plane (where walk-off occured), the pump was polarized  along the y-direction, and the BSV along the x-direction.  
The pump radiation was rejected with dichroic mirrors DM and the BSV was collected with lens $L_1$ (focal length $15$ cm), its Fourier plane shown by a dotted line (F). At this position we placed a spectrometer to measure the  wavelength spectrum of the BSV. It was centered at 1600 nm, and under collinear phase matching it had a bandwidth of $\Delta \lambda = 360$ nm (Fig. \ref{fig:Experimental-setup} (a)). The angular spectrum, measured with a camera at the same position, is shown in Fig. \ref{fig:Experimental-setup} (b). In all further measurements, to get rid of the walk-off effect we reduced the angular spectrum in the y-z plane to $10$ mrad by placing a 1.6 mm slit in plane F along the x-direction.

The BSV beam was then sent into the autocorrelator through a 50:50 beam splitter and the time delay was introduced in one arm by retroreflector $R_1$. In the other arm, a broadband half-wave plate HWP changed the polarisation from `x' to `y'. This modification eliminated the first-order interference at the output of the second beamsplitter $BS_2$, where the two beams were combined. The spatial displacement was generated by moving $BS_2$ transversely to the beam propagation direction. Noteworthy, lens $L_1$ imaged the output face of the BBO crystal on $BS_2$ with an $M=8$ magnification, which we took into account while measuring the width of the spatial correlation. 

Further, lens $L_2$ imaged $BS_2$ on the SFG crystal, which was a $500$ $\mu$m thick z-cut $\alpha$-quartz. The SFG was non-phasematched~\cite{kopylov2019study}, which made it broadband and let all frequency components of BSV contribute ~\cite{brambilla2012disclosing}. The dependence of the SFG signal on the input polarization (Fig.~\ref{fig:Experimental-setup} (c)) is typical for type-II SFG, and it originates from the structure of the nonlinear tensor of quartz. As a result, there was no SFG from either arm of the autocorrelator, which had `x' ($0^o$) or `y' ($90^o$) polarization, but only from both. The SFG radiation was then collected and sent, after polarizer P, to a spectrometer. The SFG spectrum obtained for a perfectly balanced autocorrelator is shown in Fig~\ref{fig:Experimental-setup} (d). The peak, known as the {\it coherent contribution}, is formed by merging photons from the same pair, and its area (red) corresponds to the real coincidences in a Hanbury Brown -- Twiss (HBT) measurement. The background is caused by photons from different pairs and its area (blue) corresponds to the `thermal' and `accidental' coincidences in a HBT setup~\cite{jedrkiewicz2011high,abram1986direct}.

\section*{Frequency and wavevector correlations}
\label{sec:Frequency-wavevector correlations}

\begin{figure}[htbt] 
	\centering
	\includegraphics[width=0.9\linewidth]{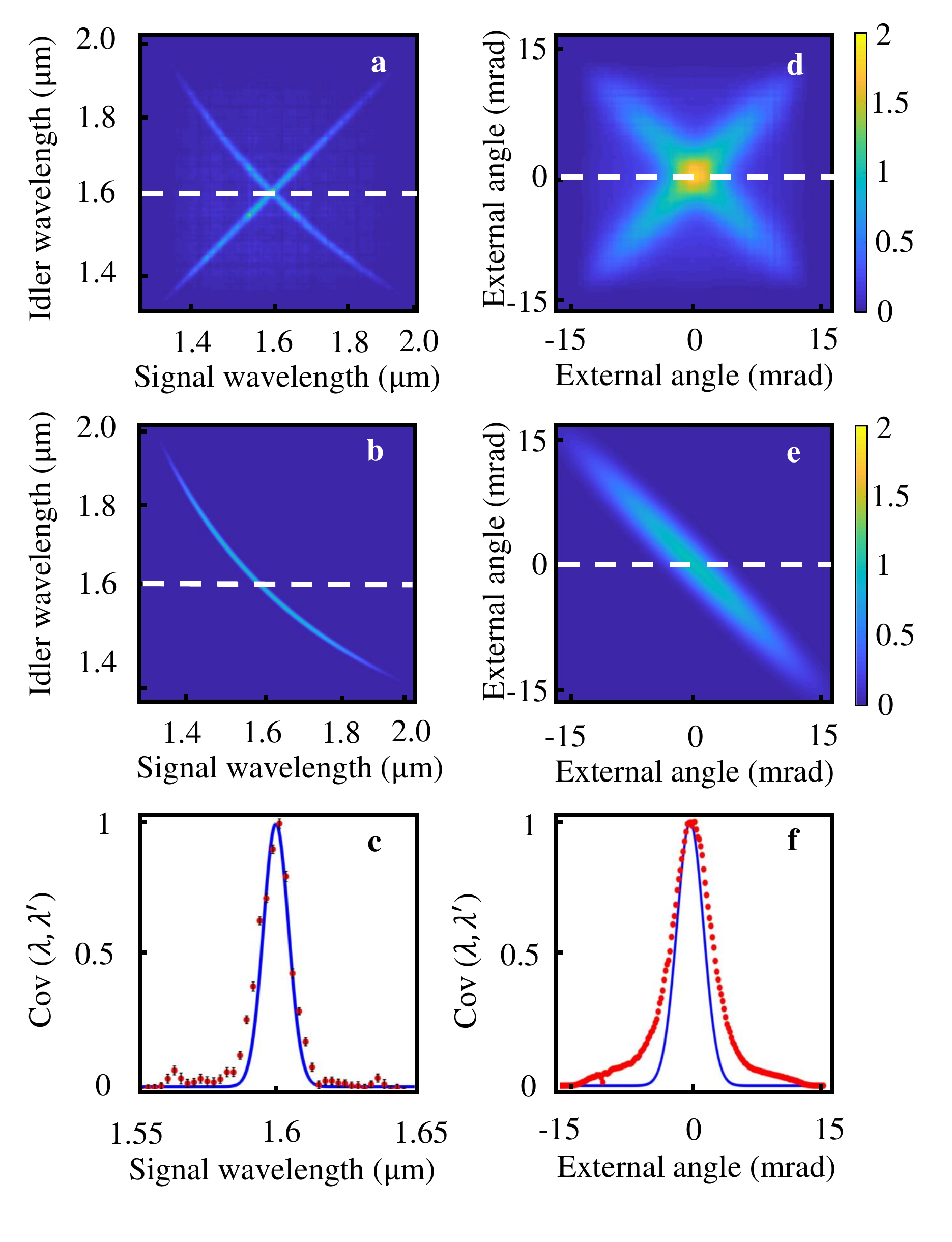}
	\caption{Intensity covariance measured versus wavelengths (a) and angles (d) and its cross-correlation part calculated for the same cases (b,e). Panels (c) and (f) show, respectively,  cross-sections of panels (a,d) (points) and panels (b,e) (lines) at $\lambda = 1.6 \:\mu$m and at $\theta = 0 ^o$ (white dashed lines), normalized to the maximum values}.  \label{fig:covariance}
\end{figure}



To find the frequency correlation width $\delta\omega$ of BSV, we obtained its spectra by placing a multimode fiber tip, connected to a spectrometer, in the Fourier plane (F in Fig~\ref{fig:Experimental-setup}). Instead of the CF $g^{(2)}(\omega,\omega')$ mentioned above, we measured the covariance~\cite{spasibko2012spectral,beltran2017orbital} of the intensity $I(\lambda)$ at different wavelengths $\lambda, \lambda'$: 
\begin{equation}
\hbox{Cov}(\lambda,\lambda')=\langle I(\lambda) I(\lambda') \rangle - \langle I(\lambda)\rangle \langle I(\lambda') \rangle.
\label{eq:covariance}
\end{equation}
This value is related to the CF $g^{(2)}(\lambda,\lambda')$ as $\hbox{Cov}(\lambda,\lambda')= \langle I(\lambda)  \rangle \langle I(\lambda')\rangle \big( g^{(2)}(\lambda,\lambda')-1 \big)$. The frequency correlation width at wavelength $\lambda$ is then $\delta\omega=2\pi c\delta\lambda/\lambda^2$, where $c$ is the speed of light and  $\delta\lambda$ is the FWHM of the covariance distribution. Figure~\ref{fig:covariance}(a) shows the covariance calculated according to Eq.~(\ref{eq:covariance}) from 4000 spectra obtained in the collinear configuration, each one acquired over 5 pulses. The shape seen along the main diagonal ($\lambda=\lambda'$) is caused by the auto-correlation of each frequency mode with itself. The shape along the anti-diagonal demonstrates the cross-correlations between signal and idler photons. The maxima follow the energy conservation condition, $\omega_p=\omega+\omega'$ ($\omega_p$ is the pump frequency), i.e., $\lambda^{-1}+\lambda'^{-1}=\hbox{const}$. The normalization is chosen such that for both cross-correlation and anti-correlation part, the maximal value is unity. The FWHM of the covariance cross-section (panel (c), red points) yields the range of correlations $\delta\lambda=12$ nm, corresponding to $\delta\omega=8.8\cdot10^{12}$ Hz. 

To calculate $\delta\omega$, we use the standard expression for the two-photon probability amplitude $F(\omega,\omega',k,k')$, where, to take into account the high-gain regime, the `sinc' function is replaced by its hyperbolic analog~ \cite{spasibko2020spectral,brambilla2004simultaneous}:                     
\begin{equation}
F(\omega,\omega',k,k')=F_{p}(\omega+\omega'-\omega_p)F_p(k+k') \frac{G}{\tilde{G}}\sinh\tilde{G} .
\label{eq:F(w,k)}
\end{equation}
Here, $F_p(\omega)$ and $F_p(k)$ are the pump Gaussian spectral and transverse wavevector amplitudes, respectively~\cite{mikhailova2008biphoton}, $\tilde{G}\equiv \sqrt{G^2-(\Delta(\omega,\omega',k,k'))^2L^2/4}$, $\Delta (\omega,\omega',k,k')$ is the wavevector mismatch and $L$ is the crystal length.
In the high-gain regime of PDC, the pump pulse duration is effectively reduced (by a factor of 3.2 for $G=10$)~\cite{sharapova2020properties} because BSV is mainly generated around the pulse peak, and we take it into account by assuming that the pump spectrum broadens 3.2 times.

The cross-correlation part of the covariance was calculated as $|F(\omega,\omega',0,0)|^2$ and plotted in Fig.~\ref{fig:covariance}(b). Its cross-section (panel (c), blue line) gives the value of $\delta\lambda=10$ nm. The somewhat larger value obtained in experiment is due to the finite resolution of the spectrometer ($7$ nm).

A similar procedure was performed to find the angular width of correlations. We replaced the spectrometer with a camera preceded by a $12$ nm bandpass filter centered at $1600$ nm. From 9000 2D frames, each one acquired over 5 pulses, we selected the central rows to pass to 1D angular spectra and calculated the intensity covariance. The result is shown in Fig.~\ref{fig:covariance} (d). As in the case of wavelengths, the angular distribution of the covariance shows the auto-correlation part (diagonal) and the cross-correlation part (anti-diagonal). Panel (e) of the same figure shows the calculated cross-correlation. Similarly to the case of wavelengths, the covariance was calculated as $|F(\omega_p/2,\omega_p/2,k,k')|^2$, with the effective narrowing of the pump beam waist by a factor of $3.2$  taken into account. Panel (f) shows the cross-sections of the experimental (red points) and the calculation (blue line) distributions. The measured angular width of correlations is $\delta \theta= 4.1 $ mrad, somewhat larger than the calculated value $3.4$ mrad. The resulting width in the transverse wavevector is $\delta k = 0.016 \: \mu$m$^{-1}$.

\section*{Spatiotemporal correlations}
\label{sec:st correlations}

We used the autocorrelator to measure the dependence of $g^{(2)}(t,t',x,x')$ on $\tau\equiv t-t'$ and $\xi\equiv x-x'$. For a fixed position of $BS_2$ we scanned the time delay in one arm with $R_1$, similar to Ref.~\cite{cutipa2020direct}. For each time delay $\tau$ between the two arms, we obtained a spectrum as shown in Fig.~\ref{fig:Experimental-setup} (d). The height of the peak strongly depends on $\tau$ and becomes zero at large delays, indicating that only photons that belong to different pairs merge into a sum-frequency photon. From the spectra, we  calculated the value of the CF, according to Ref.~\cite{kopylov2020spectral}, as $g^{(2)}(\tau)=2(B(\tau)/B_{max}+R(\tau)/B(\tau))$, where $R(\tau)$ is the integral spectrum of the peak (red area in Fig.~\ref{fig:Experimental-setup} (d)), $B(\tau)$ the integral spectrum of the background (blue area in Fig.~\ref{fig:Experimental-setup} (d)), and $B_{max}$ its maximal value, which is achieved at $\tau=0$.

\begin{figure}[btth]
	\centering
	\includegraphics[width=0.85\linewidth]{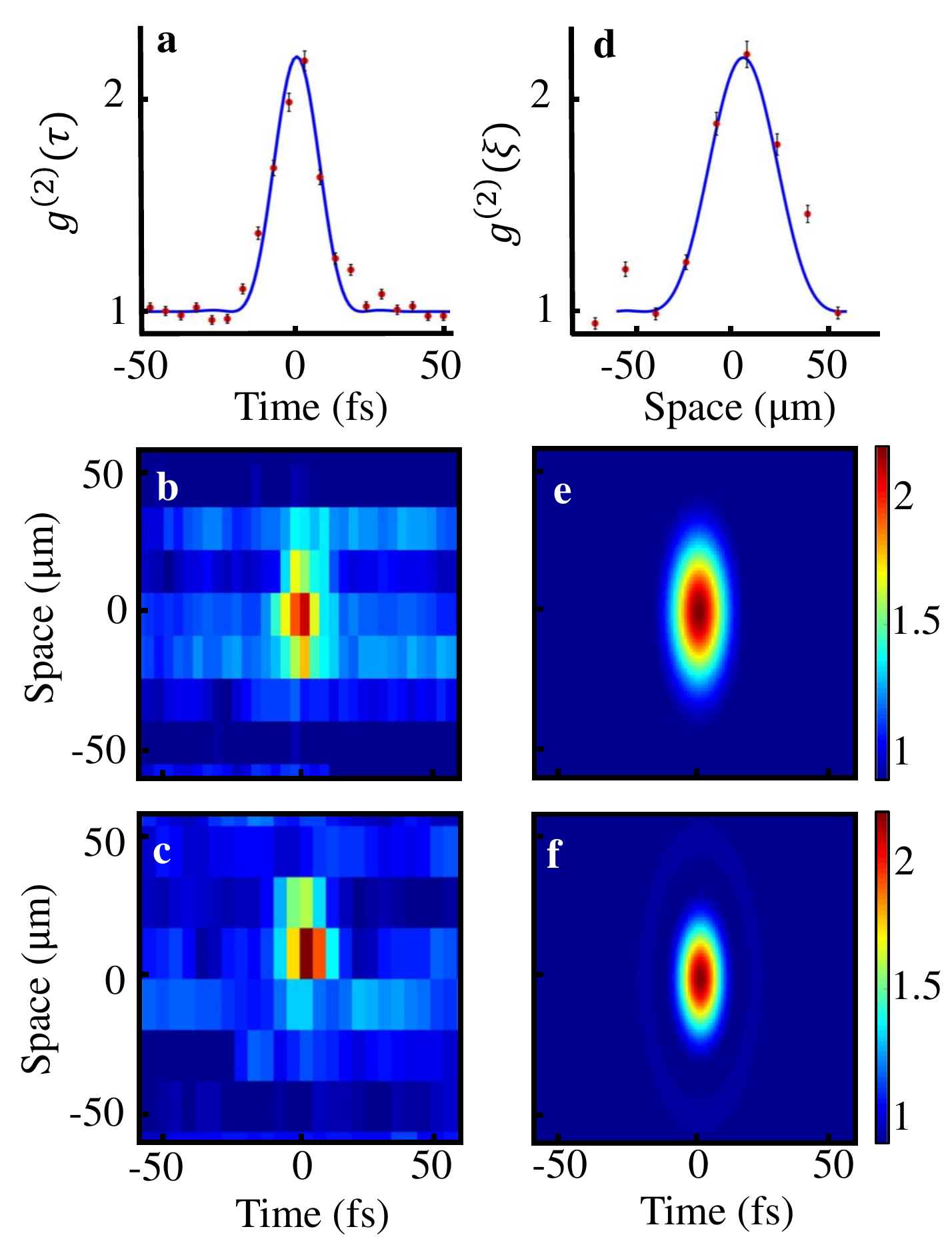}
	\caption{Second-order CF $g^{(2)}(\tau,\xi)$ measured (b,c) and calculated (e,f) for the collinear (b,e) and non-collinear (c,f) PDC. Panels (a) and (b) show the profiles of panels (b) (points) and (e) (line) at $\xi=0$ and $\tau=0$, respectively.} \label{fig:exp_result_time_space}
\end{figure}
This is how we obtained the 1-dimensional plot shown in Fig.\ref{fig:exp_result_time_space} (a) for $\xi = 0$. The correlation time $\delta\tau$ is equal to its full width at half maximum (FWHM). We repeated this measurement for each spatial displacement $\xi$ made with $BS_2$. Importantly, each spatial displacement generated an extra time delay, which we compensated by retroreflector $R_1$. By stacking together $g^{(2)}(\tau)$ distributions at different $\xi$, we obtained the two-dimensional distribution $g^{(2)}(\tau,\xi)$  (Fig.~\ref{fig:exp_result_time_space}) for both collinear (b) and non-collinear (c) PDC.  Panels (e) and (f) show the corresponding theoretical distributions, calculated as squared two-dimensional Fourier transforms of Eq.(\ref{eq:F(w,k)}). The distributions for the non-collinear configuration are narrower due to a broader wavelength-angular spectrum. In agreement with the theoretical calculation, the correlation time and distance for the collinear case are, respectively, $18\pm 2$ fs and $45 \pm 6$ $\mu$m, and for the non-collinear case, $12\pm 2$ fs and $33 \pm 6$ $\mu$m.  The peak value $g^{(2)}(0,0)$ obtained is $2.2$ and not $3$, as expected, due to the group-velocity dispersion in the optical elements. Panel (d) shows the space profile of the CF at $\tau=0$ for collinear PDC: the points and the line are the cross-sections of panels (b) and (e), respectively.

\begin{figure}[h]
	\centering
	\includegraphics[width=0.9\linewidth]{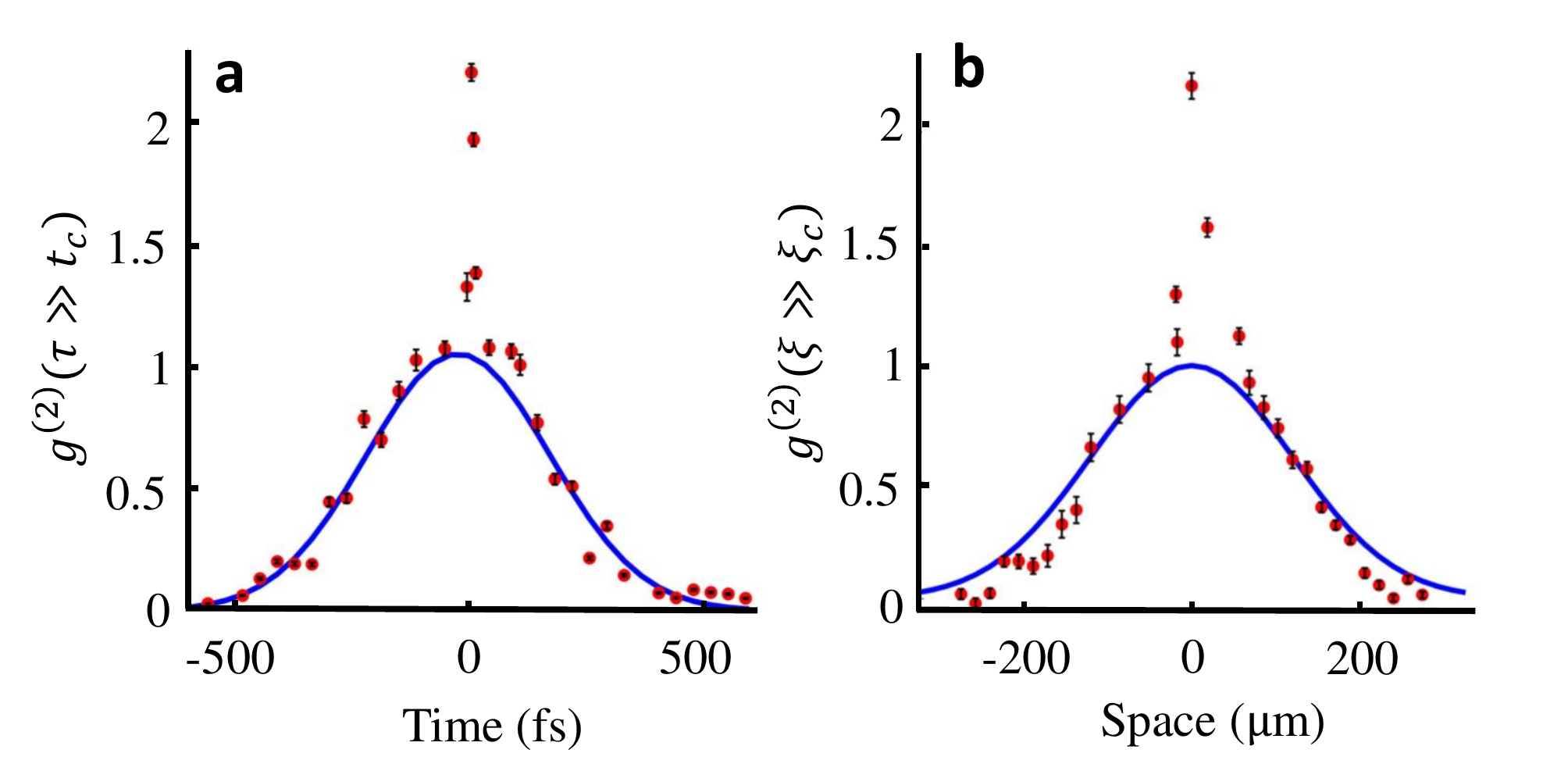}
	\caption{The CF $g^{(2)}(\tau)$ (a) and $g^{(2)}(\xi)$ (b) measured over larger time and space scales. The pedestals follow Gaussian distributions (blue lines) given by the BSV pulse duration (a) and beam waist (b).} \label{fig:large_spatial_time}
\end{figure}

Figure \ref{fig:exp_result_time_space} shows the CF $g^{(2)}(\tau,\xi)$ distributions only in the vicinity of its maximum at $\tau=\xi=0$. We also measured its behavior at a larger scale, with the results shown in Fig.~\ref{fig:large_spatial_time} (a,b). The background follows a Gaussian distribution (blue line) whose widths in time and space are determined, respectively, by the BSV pulse duration and the beam waist. In agreement with the above-mentioned effect of the high gain, both widths are indeed $3.2$ times smaller than those for the pump \cite{brida2010,allevi2014coherence,sharapova2020properties}.

Finally, we calculate the time-frequency and space-wavevector products.
For collinear degenerate PDC, we get $\delta\tau \delta \omega = 0.16$ and $\delta\xi\delta k = 0.72$. Both values are far below the Mancini (or Fourier) bound $2\pi$.  We note that the second condition, $\delta\xi\delta k \ll 2\pi$, was also demonstrated recently for BSV obtained through four-wave mixing, using noise reduction measurement~\cite{kumar2021einstein}.

In conclusion, we measured the the second-order CF for BSV as a function of space and time. We obtained the temporal and spatial correlation widths of $\delta\tau=18$ fs and $\delta\xi=45\,\mu$m  for the collinear degenerate configuration and $\delta\tau=12$ fs and $\delta\xi=33\,\mu$m for the non-collinear non-degenerate configuration, respectively. These values define the resolution for two-photon spectroscopy with BSV. 
The high temporal and spatial resolution is useful for analyzing complex samples and extracting information about their energy levels. In addition, we found the correlation widths of BSV in frequency and transverse wavevector, $\delta\omega=8.8\cdot10^{12}$  Hz, $\delta k= 0.016 \: \mu$m$^{-1}$, respectively. As a result, we have demonstrated that BSV enables  simultaneously high resolution in time and frequency, and space and wavevector. 

We thank Tom\'as Santiago-Cruz for the help in experiment and Denis Kopylov for helpful discussions.

\section*{Funding.}
P. Cutipa acknowledges the funding provided by the National Agency for Research and Development (ANID) / DOCTORADO BECAS CHILE/2017 - 72180453.

\section*{Disclosures.} The authors declare no conflicts of interest. 



\printbibliography 
 

\end{document}